\documentclass[12pt]{article}

\usepackage{epsfig}
\usepackage{axodraw}

\def \lsim{\mathrel{\vcenter
     {\hbox{$<$}\nointerlineskip\hbox{$\sim$}}}}

\newcommand{\beq}{\begin{equation}}
\newcommand{\eeq}{\end{equation}}
\newcommand{\beqa}{\begin{eqnarray}}
\newcommand{\eeqa}{\end{eqnarray}}
\newcommand{\beqar}{\begin{eqnarray*}}
\newcommand{\eeqar}{\end{eqnarray*}}

\begin{document}

\thispagestyle{empty}

\hfill{}

\hfill{}

\vspace{32pt}

\begin{center} 
\textbf{\Large 
Cosmic Rays and New Physics at the TeV: the Neutrino-Nucleon 
Cross Section} 

\vspace{40pt}

M. Masip

\vspace{12pt}

\textit{Centro Andaluz de F\'\i sica de Part\'\i culas Elementales
(CAFPE)\\ and\\ Departamento de F{\'\i}sica Te\'orica y del
Cosmos}\\ \textit{Universidad de Granada, E-18071, Granada, Spain}\\
\vspace{6pt}
\texttt{masip@ugr.es}
\end{center}

\vspace{40pt}

\begin{abstract} 

Ultrahigh energy neutrinos can be used to explore the physics
at the TeV scale. We study the neutrino-nucleon cross section 
in models
with extra dimensions and the fundamental scale at the TeV. 
In particular, we discuss the production of 
string resonances and the gravitational interactions 
(multigraviton exchange and production of microscopic 
black holes) in these models.
We show that the new TeV physics 
could give observable signals in
horizontal air showers and neutrino telescopes.

\end{abstract}

\setcounter{footnote}{0}

\newpage

\section{Introduction}

We observe extensive air showers produced when a cosmic ray from
outer space hits a nucleon in the upper atmosphere.
The energy of the particle starting the shower can
be very large, with observed events of up to $10^{12}$ GeV. In
particular, there are events above the so called
GZK cutoff energy $E_{\rm GZK}\approx 5\times 10^{19}$~eV. 
These events present a problem because, although the profile
of the shower is consistent with a primary proton, the process
\beq
p + \gamma_{2.7{\rm K}} \to \Delta^+\to p+\pi^0 \ (n+ \pi^+)
\eeq
is very effective on reducing the energy of a proton 
propagating in the cosmic background. Since in principle there are 
no near sources of such energetic particles, 
these cosmic rays should not be protons. One could then especulate 
if they are neutrinos. Neutrinos can come from a very far source
with no loss of energy and following straight trajectories, which 
would explain the observation of doublets and triplets of events. 
The problem with neutrinos is that their standard 
cross section (c.s.) with the atmosphere is five orders of magnitude 
too small. A first motivation to study the neutrino-nucleon c.s.
is then: Can new TeV physics increase this c.s. up to hadronic size? 

A second motivation would go the
other way around. There are experiments designed to measure the
standard model (SM) neutrino-nucleon c.s. at ultrahigh
energies. The question is: Can new TeV physics give 
observable signals in horizontal air showers or
neutrino telescopes? Note that 
since they are weakly interacting the relative effect of 
new physics on neutrino interactions is going to be much 
larger than in proton or charged lepton interactions. 

Let us make a naive estimate of the impact of new physics
on the cross section. The $\nu$-$q$ amplitude 
is mediated in the SM by a $Z$ boson in the $t$ channel: 
${\cal A}_{\rm SM}(s,t)\propto 
g^2 s/(t-M^2_Z)$, where $s$ and $t$ are the 
usual Mandelstam variables ($t=-q^2=-\frac{1}{2}(1-\cos\theta)$).
In the limit $s\gg M_Z^2$ the c.s. 
\beq
{\sigma}_{\rm SM}(s)=\frac{1}{16\pi}
\frac{1}{s^2}
\int^0_{-s} {\rm d}t\ |{\cal A}_{\rm SM}|^2 
\propto \frac{g^4}{M^2_Z}-\frac{g^4}{s+M^2_Z}
\eeq
is dominated by $t$ of 
order $M_Z^2$, and becomes just ${\sigma}_{\rm SM}(s)\propto g^4/M^2_Z$. 
This simple result tells us that even at large center of mass
energies the c.s. is very sensitive to the mass of the
exchanged particle. It explains why the c.s. would be
much larger if the exchanged particle were a gluon or a photon,
and also that a $Z'$ boson would introduce only a small correction 
to the SM result.

The presence of extra dimensions would have a more promising 
impact. A graviton-mediated 
Born amplitude will present two main features. 
First, the spin 2 of the
intermediate field gives amplitudes growing like $s^2$, versus
just $s$ for the spin 1 $Z$ boson. Second, one has to sum the 
contributions of the infinite tower of KK gravitons. Actually, 
this gives a divergence for more than one extra dimension.
Although both effects push the cross section in the right 
direction, they both imply the presence of an ultraviolet 
cutoff $\Lambda$ above which the
model is not consistent. It is then easy to conclude that at
the cutoff the c.s.  
$\sigma_{4+n}\approx {1/\Lambda^2}$ does not seem large 
enough.

String theory provides another scenario with an infinite tower 
of higher spin fields (the string excitations) which, in addition, 
does not require an ultraviolet cutoff. In the brane world 
picture matter and gauge fields correspond to the zero modes 
of open strings, whereas the graviton is the massless
mode of a closed string. A four fermion amplitude will include
diagrams exchanging open and closed strings, with the later subleading
in the string coupling $g$ (the exchange of a closed string 
can be also seen as the one-loop exchange of two open 
strings). We would expect: 

\noindent {\it (i)} At $s\lsim M_S^2$ the diagram with exchange of
an open string dominates the amplitude, giving at $s\approx 0$
a $Z$ boson in the $t$ 
channel. The diagram with an intermediate closed
string gives at low energy the gravitational interactions. 

\noindent {\it (ii)} At $s\approx M^2_S$ both types of diagrams
give string resonances (Regge excitations). Closed string 
excitations, however, couple weaker
($\sim g^2$) to fermions. 

\noindent {\it (iii)} At $s\gg M_S^2$ both 
diagrams give the usual soft behaviour of the string in the
ultraviolet: the amplitudes go to zero exponentially
at fixed angle ($t/s$ fixed) and like a power law in the Regge
limit ($t$ fixed). Essentially, the string
amplitude goes to zero everywhere except forward 
($-t\lsim M_S^2$), where only survives the
contribution of the massles mode of the intermediate string.
In this regime the exchange of open and closed string 
gives, respectively, 
\beq
{\cal A}(s,t)\approx g^2 {s \over t}
\eeq
and
\beq
{\cal A}(s,t)\approx g^4 {s^2 \over M_S^{n+2}} 
\int {{\rm d}^n\;q_T\over t-q_T^2}
\eeq
Although the amplitude mediated by a closed string is 
subleading in $g$, it grows faster
with $s$. As the center of mass energy increases the string
amplitude is dominated by the long distance (small $t$) contributions
of the higher dimensional graviton.

In the next sections we will evaluate the contributions to the
$\nu$-$N$ c.s. from the production of open string excitations and 
from graviton exchange.
We will go beyond the Born level and use the eikonal 
approximation to resumate the dominant long distance graviton 
contributions. At shorter distances (but still larger than $1/M_S^2$) 
nonlinear effects become important. There we will 
estimate the cross section to produce a microscopic black hole.

\section{String excitations}

We will consider a simple brane model where matter and gauge
fields are the zero modes of open strings with both ends attached to
a set of $N$ 4-dimensional branes. These D3-branes would be
sitting at a fixed point of a higher dimensional bulk.
Such a framework would
define a model with $U(N)$ gauge symmetry and N=4 SUSY, but we
will asume that an orbifold projection eliminates the extra 
symmetry of the massless modes leaving just the SM.

The four fermion amplitude corresponding to
the exchange of an open string is then very simple:
\beqa
{\cal A}(1,2,3,4)&=&
   {g^2} {{\cal S}(s,t)} {F^{1243}(s,t,u)} 
{{\rm Tr}[t^1t^2t^4t^3+t^3t^4t^2t^1]}
\nonumber\\
&+&{g^2} {{\cal S}(s,u)} {F^{1234}(s,u,t)} 
{{\rm Tr}[t^1t^2t^3t^4+t^4t^3t^2t^1]}
\nonumber\\
&+&{g^2} {{\cal S}(t,u)}{ F^{1324}(t,u,s)} 
{{\rm Tr}[t^1t^3t^2t^4+t^4t^2t^3t^1]\;.}
\eeqa
In this expression $g$ is the gauge coupling, 
${\cal S}(s,t)=\Gamma(1-\alpha's)\Gamma(1-\alpha't)/
\Gamma(1-\alpha's-\alpha't)$ with $\alpha'=M_S^{-2}$
is the Veneziano factor ($M_S$ is the string scale),  
the factors $F^{abcd}$ depend 
on the helicity of the external fermions,
and the Chan-Paton traces describe the gauge numbers  
of the fermions ($t^a$ are representation
matrices of $U(N)$).

Let us consider the process $\nu_L u_L\rightarrow \nu_L u_L$.
We find that if the Chan-Paton traces satisfy
$T_{1243}-T_{1324}=-1/10$ and $T_{1234}=T_{1324}\equiv -a/10$
then at low energies 
the string amplitude reproduces the SM result: a massless
$Z$ boson (it gets mass only through the Higgs mechanism) 
in the $t$ channel with  $s^2_W=3/8$ (as our model has
only one gauge coupling). In particular, for $a=0$ our 
string amplitude reads
\beq
{\cal A}(s,t)=\frac{2}{5}g^2\frac{s}{t} {\cal S}(s,t)\;.
\eeq
This amplitude has poles at $s=nM_S^2$, corresponding to
the exchange of string resonances in the $s$ channel. Near the 
$n$ pole the amplitude is
\beq
{\cal A}_n\approx \frac{2}{5}\frac{g^2}{s-nM^2_S}
\frac{nM^4_S}{t}\frac{(t/M^2_S)(t/M^2_S+1)\cdots(t/M^2_S+n-1)}{(n-1)!}
\eeq
The residue is a polinomial of order $n-1$ in $t$, indicating that
the spin of the resonances in this mass level goes up to $J=n-1$.
To separate the contribution of each resonance we express the
amplitude in terms of the scattering angle and parametrice it
in terms of rotation matrix elements:
\beq
{\cal A}_n=\frac{2}{5}g^2\;\frac{nM^2_S}{s-nM^2_S}\;\sum^{n-1}_{J=0}
{\alpha^J_n}\ d^J_{00}(\theta)\;.
\eeq
The coefficient $\alpha^J_n$ gives the contribution to 
${\cal A}(\nu_L,u_L\rightarrow \nu_L,u_L)$ of the resonance
with mass $\sqrt{n}M_S$ and spin $J$. For example, at the 
first mass level
we obtain a scalar resonance with $\alpha^0_1=1$, 
at $s=2M_S^2$ there is a single vector resonance
with $\alpha^1_2=1$, whereas at 
$s=3M_S^2$ there are modes of spin 
$J=2$ ($\alpha^2_3=3/4$) and $J=0$ ($\alpha^0_3=1/4$).
We find an interesting sum rule for the $\alpha^J_n$ 
coefficients: $\sum_{J=0}^{n-1} \alpha^J_n=1$, independent of
the mass level $n$. 

We would like to emphasize that the
presence of leptoquarks (color and isospin triplets) 
mediating the $\nu$-$q$ amplitude in
the $s$ channel is not a peculiarity of our toy model but
a generic result. Due to the $s$-$t$ duality of the string
amplitude they will appear in any model that has the SM as low
energy limit.

From the elastic $\nu_L$-$u_L$ amplitude we can deduce the
c.s. to produce resonances in the narrow width 
approximation and estimate $\nu_L u_L\rightarrow anything$ from 
$\nu_L u_L\rightarrow X^J_n \rightarrow anything$. The
estimate will be good for energies up to 
$ n(2/5) g_0^2/(4\pi) \approx 1$.
Above $M_X\approx 7 M_S$ ($n=50$) the width of the resonances 
is similar to the energy separation
with the next mass level and the amplitude is dominated by
interference effects.

We obtain that the partial width of the spin $J$ resonance at the
$n$ mass level is 
\beq
\Gamma(X^J_n\to\nu_Lu_L)\equiv
 {\Gamma^J_n}=\frac{2}{5}
\frac{g^2|{\alpha^J_n}|}{16\pi}\frac{\sqrt{n}M_S}{2J+1}\;.
\eeq
Taking 
\beq
\frac{M\Gamma}{|\hat{s}-M^2+i\Gamma M|^2}\to\pi\delta(\hat{s}-M^2)
\eeq
we find
\beq
\sigma(\nu_Lu_L\to {X^J_n}\to\mbox{all}) \equiv \sigma^J_n(\nu_Lu_L)
=\frac{4\pi^2\Gamma^J_n}
{\sqrt{n}M_S}(2J+1)\delta(\hat{s}-nM^2_S)\;,
\eeq
and the cross section to create in the collision any resonance at the
$n$ mass level is just
\beq
\sigma_n(\nu_Lu_L)=\sum^{n-1}_{J=0}\sigma^J_n=
\frac{\pi g^2}{4}\;{2\over 5}\;\delta(\hat{s}-nM^2_S)\;,
\eeq
independent of the mass level.

We repeat the procedure for the rest of partons in the
nucleon and plot in Fig.~1 the total $\nu$-$N$ c.s. 
to create string resonances for a particular choice of
Chan-Paton traces.

\vspace{0.3cm}

\begin{figure}[ht]
\begin{center}\leavevmode  %
\epsfxsize=10.cm
\epsfbox{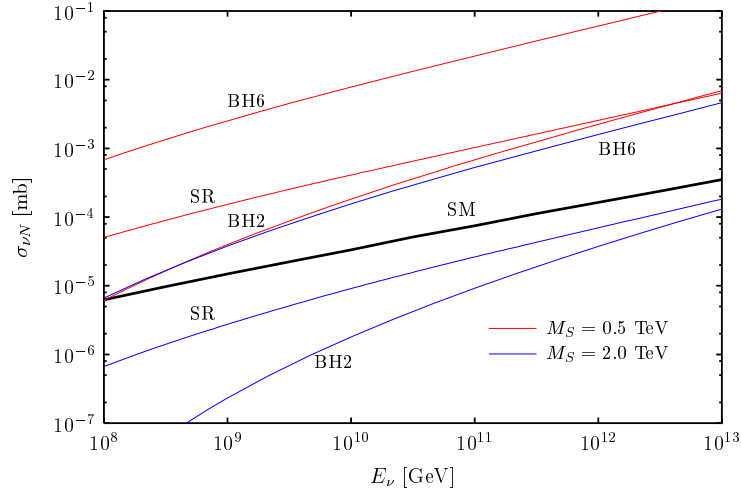}
\caption{ Cross section to produce string excitations (SR) and black holes
in 2 or 6 extra dimensions (BH2, BH6). We include the SM cross section.
\label{Fig. 1}}
\end{center}
\end{figure}

\section{Graviton exchange: eikonal approximation}

Although matter and gauge fields are attached to a 4-dimensional
brane, gravity propagates in the 10 dimensions. We will asume 
that $n$ of the 6 extra compact dimensions are large and gravity 
is D-dimensional ($D=4+n$) at the distances of interest.
The higher dimensional
Newton's constant is usually defined as 
\beq
G_D= V_n G_N \equiv {(2\pi)^{n-1}\over M_D^{n+2}}\;,
\eeq
where $V_n$ is the volume of the compact space, 
$G_N$ is the 4-dimensional constant, and $M_D$ the higher 
dimensional Planck scale. Obviously, $M_D$ is related to the
string scale $M_S$. From the low-energy limit of the 
closed string amplitude we obtain
\beq
G_N = {g^4\over 64\pi^2}\;
{1\over M_S^{n+2} R^n}\quad \Rightarrow\quad
M_D^{n+2}={1\over 2 \pi \alpha^2} M_S^{n+2}\;.
\eeq
For $n=2$ $M_D=3.5M_S$, whereas for $n=6$ $M_D=1.9M_S$.

As explained in the introduction, at transplanckian energies
the $\nu$-$q$ amplitude will be dominated by 
forward (long distance) graviton-mediated contributions.
This is precisely the regime where we can use the eikonal
approximation, that resumates all the 
ladder and cross-ladder contributions. Essentially, it is the
exponentiation of the Born amplitude in impact parameter
space:
\beqa
{\cal A}_{eik}(s,t)&=&{2 s\over i}
\int d^2b\; e^{i\mathbf{q}\cdot\mathbf{b}}\;
\left(e^{i\chi (s,b)}-1\right) \nonumber \\
&=&{4\pi s\over i}\int db\;bJ_0(b q)\;\left(e^{i\chi (s,b)}-1\right)\;,
\eeqa
where $\chi (s,b)$ is the eikonal phase fixing the amplitude
and $\mathbf{b}$ spans the (2-dimensional) impact parameter space.
The Born amplitude 
corresponds to the limit of small $\chi (s,b)$:
\beq
{\cal A}_{Born}(s,t)=
{4\pi s\over i}\int db\; bJ_0(b q)\; i\chi (s,b)\;.
\eeq
We can then deduce the eikonal phase from the Fourier transform
to impact parameter space of the Born amplitude:
\beq
\chi(s,b)={i\over 2s}\int {d^2
q\over(2\pi)^2}\; e^{i{\bf q}\cdot{\bf b}}\;i{\cal A}_{Born}\;.
\eeq
In our case the ${\cal A}_{Born}$ comes from the exchange in the $t$ 
channel of a higher dimensional graviton:
\beq
i{\cal A}_{Born}=i{s^2 \over M_D^{n+2}} 
\int {{\rm d}^n\;q_T\over t-q_T^2}=
i\pi^{n/2}\Gamma\left(1-{n\over 2}\right)
{s^2\over M_D^{2+n}} (-t)^{{n\over 2}-1}\;,
\eeq
where the integral over momentum along the extra dimensional $q_T$
(equivalent to the sum over KK modes) gives an ultraviolet 
contribution that we have regularized using dimensional 
regularization. The {\it magic} of the eikonal approximation is
that it will be well defined although we obtain it from an ultraviolet
dependent Born amplitude: the contributions from large $q_T$ introduce
corrections to the phase $\chi (s,b)$ only at small $b$ 
($\approx 1/q_T$), but this small $b$ region (see Eq.~(15)) gives 
no sizeable contribution to the eikonal amplitude.

We obtain that the eikonal phase is 
\beq
\chi(s,b)=(b_c/ b)^n\;, \quad b_c^n=
{(4\pi)^{{n\over 2}-1}\over 2}\Gamma\left({n\over 2}\right)
{s\over M_D^{2+n}}\;.
\eeq
We note that $\chi(s,b)$ introduces a new scale, $q_c=1/b_c$, 
that sets the
size of the total cross section. From the optical theorem we 
deduce
\beq
\sigma_{el}={{\rm Im}{\cal A}_{eik}(s,0)\over s}
=2\pi b_c^2\;\Gamma\left(1-{2\over n}\right)\cos{\pi\over n}
\sim s^{2/n}\;.
\eeq

${\cal A}_{eik}(s,t)$ can be evaluated numerically from Eq.~(15) 
for any values
of $s$ and $t$. We can also obtain approximate expressions in the
limits of small and large $q\equiv \sqrt{-t}$. For $q\ll q_c$ 
we get an expression analogous to the 
Born amplitude but with an effective cutoff of ${\cal O}(q_c)$. 
In particular, at $t=0$ the amplitude goes to
\beq
{\cal A}_{eik}(s,0)=2\pi i\; s b_c^2\; \Gamma\left(1-{2\over n}
\right)e^{-i\pi\over n}\;.
\eeq
For $q\gg q_c$ the integral over impact parameter space is dominated
by a saddle poing at $b_s = b_c(qb_c/n)^{-1/(n+1)}\ll b_c$, resulting
\beq
{\cal A}_{eik}(s,q)=Z_n\left({s\over
qM_D}\right)^{n+2\over n+1}\;,
\eeq
with $Z_n$ a complex phase.

We give in Fig.~2 the $\nu$-$N$ eikonal cross section in terms of
the fraction of energy $y=(E_\nu-E'_\nu)/E_\nu$ lost by the incident 
neutrino. For low values of $n$ the neutrino would interact with
the atmosphere, but losing only a small fraction of its energy
(the cross section could be very large but {\it soft}).

\vspace{0.3cm}

\begin{figure}[ht]
\begin{center}\leavevmode  %
\epsfxsize=10.cm
\epsfbox{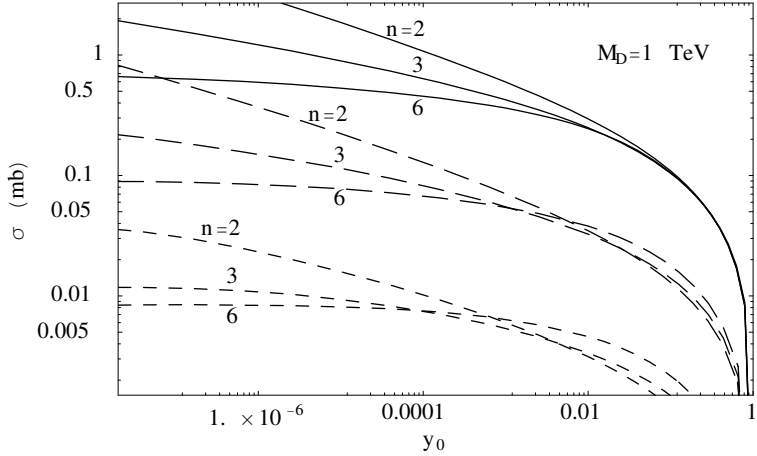}
\caption{ Elastic cross section vs.\ minimum fraction of energy  lost
by the neutrino for $E_{\nu}=10^{14}$ GeV (solid), $E_{\nu}=10^{12}$ GeV 
(long dashes) and $E_{\nu}=10^{10}$ GeV (short dashes).
\label{Fig. 2}}
\end{center}
\end{figure}

\section{Black hole production}

The eikonal approximation provides an acceptable description 
of the scattering as far the dominant
impact parameters (given by the position of the saddle
poing $b_s$) are larger than the Schwarzschild radius $R_S$ of
the system: 
\beq
R_S=\left({2^n\pi^{n-3\over 2}\Gamma\left({n+3\over 2}\right)
\over n+2}\right)^{1\over n+1}
\left({s\over M_D^{2n+4}}\right)^{1\over 2(n+1)}\;.
\eeq
Diagramatically, at $b\lsim R_S$ one finds
that $H$ diagrams (non-linear effects of gravity) become
as important as the eikonal ladder diagrams.
In this regime the system would collapse into a microscopic
black hole with an approximate c.s.
\beq
\sigma(s)=\int_{M^2/s}^{1} dx\;
\left(\sum_i f_i(x,\mu) \right)\;
\pi R_S^2\;.
\eeq
This estimate would be reduced by graviton emision during the collapse
but enhanced by the fact that a black hole acts as a somehow
larger scatterer, so it should not be off by any large factors.
Note also that the scale in the parton distribution functions
must be $\mu=R_S^{-1}$ and not the black hole mass. The process 
becomes softer for larger center of mass energy and black
hole mass. Actually, for cross sections of order $1/m_p^2$ 
the neutrino would not see partons at all, it  would 
interact coherently with the whole proton. We give in Fig.~1
the c.s. for black hole production.

Comparing this process with the 
eikonal scattering 
we find that the energy transfer from an ultrahigh energy 
neutrino to the
atmosphere would be dominated by hard processes: 

\noindent {\it (i)} For a given flux of neutrinos of fixed 
energy, the total energy deposited in the atmosphere via
small $y$ scatterings will be smaller than through the 
(less frequent) events with large $y$ or black hole formation.

\noindent {\it (ii)} In addition, for neutrino fluxes 
$J(E)\sim E^\alpha$ with $\alpha\lsim -2$ 
a shower event of given energy would
more likely come from a neutrino of similar energy than from
a more energetic neutrino that lost only a fraction of its energy.

To conclude, the elastic exchange of higher dimensional gravitons
or the production of string excitations or microscopic
black holes can not explain the cosmic ray events above the GZK limit.
Contrary to some claims, 
the last two processes are similar in size (see Fig.~1). 
Black hole production does not dominate (especially for low values
of $n$) because the center of mass energy at the parton level is never
too large: the increasing number of partons at small $x$
favors the production of light black holes, with masses around
the fundamental scale.
All these processes, nevertheless, could give deviations to the
signals expected in horizontal
air showers and neutrino telescopes. 

\section*{Acknowledgements}

I would like to thank  
Fernando Cornet, Roberto Emparan, Jos\'e Ignacio Illana and
Riccardo Rattazzi for their collaboration, the organizers of
{\it Quarks 2002} for their kind hospitality, and the Universidad
de Granada for finantial support to attend the meeting.

\end{document}